\documentclass[a4paper]{article}

\usepackage{INTERSPEECH2020}

\usepackage{multirow}
\usepackage{hyperref}

\usepackage{epstopdf}

\DeclareMathOperator{\sign}{sign}
\DeclareMathOperator*{\argmax}{arg\,max}
\DeclareMathOperator*{\argmin}{arg\,min}

\DeclareGraphicsExtensions{.eps,.pdf}

\title{Baseline System of Voice Conversion Challenge 2020 with Cyclic Variational Autoencoder and Parallel WaveGAN}
\name{Patrick Lumban Tobing$^1$, Yi-Chiao Wu$^1$, and Tomoki Toda $^2$}
\address{
  $^1$Graduate School of Information Science, Nagoya University, Japan\\
  $^2$Information Technology Center, Nagoya University, Japan}
\email{\{patrick.lumbantobing, yichiao.wu\}@g.sp.m.is.nagoya-u.ac.jp, tomoki@icts.nagoya-u.ac.jp}

\begin{document}

\maketitle
\begin{abstract}
    In this paper, we present a description of the baseline system of Voice Conversion Challenge (VCC) 2020
    with a cyclic variational autoencoder (CycleVAE) and Parallel WaveGAN (PWG), i.e., CycleVAEPWG.
    CycleVAE is a nonparallel VAE-based voice conversion that utilizes converted acoustic features
    to consider cyclically reconstructed spectra during optimization. On the other hand, PWG is a non-autoregressive neural
    vocoder that is based on a generative adversarial network for a high-quality and fast waveform generator.
    In practice, the CycleVAEPWG system can be straightforwardly developed with the VCC 2020 dataset using a unified model for
    both Task 1 (intralingual) and Task 2 (cross-lingual), where our open-source implementation is available at
    \footnotesize{\url{https://github.com/bigpon/vcc20_baseline_cyclevae}}. \normalsize{The results of VCC 2020 have demonstrated that the
    CycleVAEPWG baseline achieves the following: 1) a mean opinion score (MOS) of 2.87 in naturalness and a speaker similarity percentage
    (Sim) of 75.37\% for Task 1, and 2) a MOS of 2.56 and a Sim of 56.46\% for Task 2, showing an approximately or nearly average score
    for naturalness and an above average score for speaker similarity.}
\end{abstract}

\section{Introduction}
\vspace*{-0.5mm}

Voice conversion (VC) \cite{Abe90} is a framework for changing the voice characteristics of a source speaker into those
of a desired target speaker while retaining the linguistic contents of the source speech. VC is very useful
for many real-world applications, such as speech database augmentation \cite{Kain98}, improvement of impaired speech \cite{Tanaka13},
entertainment purposes \cite{Kobayashi18a}, expressive speech synthesis \cite{Turk10}, and improvement of speaker verification systems
\cite{Todisco19}. To support the research and development of VC techniques, during the past six years, three VC challenges\footnotemark \hspace{0.25em}have been carried out,
i.e., VC challenge (VCC) 2016 \cite{Toda16}, VCC 2018 \cite{Jaime18}, and VCC 2020 \cite{vcc2020}. In this paper,
we provide a description for a baseline system in VCC 2020, where our goal is to provide a straightforward
baseline implementation using freely available software that can be developed even with only the VCC 2020 dataset.
\footnotetext{\scriptsize {\url{http://vc-challenge.org/}}}

In performing VC, the changes in characteristics can be affected by mainly two factors, namely, voice timbre and
prosody. Voice timbre is related to the spectral characteristics of the vocal tract during phonation,
whereas prosody is related to suprasegmental factors, such as pitch/intonation, duration, and stress.
To develop a clear-cut implementation, in this work, we keep the duration of the speech unchanged
and use a transformation of fundamental frequency (F0) to change pitch characteristics. On the other hand, we
focus on the VC development using a mapping model for spectral features, such as
mel-cepstrum parameters \cite{Tokuda94} obtained from the vocal tract spectral envelope.

Spectral modeling in VC can be developed in two ways, i.e., the parallel method using paired utterances
between speakers and the nonparallel method without using paired utterances. Many methods have been
developed for parallel spectral modeling, such as the codebook-based method \cite{Abe90}, methods using the Gaussian
mixture model \cite{Stylianou98,Toda07}, methods using dynamic kernel partial least squares regression \cite{Helander11},
and the exemplar-based method \cite{Wu14}. The nonparallel spectral modeling approach has also been
adopted in the recent years for the frequency warping method \cite{Erro10}, restricted Boltzmann machine \cite{Nakashika16},
generative adversarial network \cite{Kameoka18}, and variational autoencoder (VAE)-based model \cite{Hsu16}, among others. In this paper,
considering the versatility of the nonparallel method, especially for cross-lingual VC, we focus on the use of
nonparallel VAE-based spectral modeling.

VAE-based VC is usually developed by assuming that the latent space contains speaker-independent
characteristics, such as phonetics, while the speaker-dependent space is handled
with a time-invariant one-hot speaker code \cite{Hsu16}. However, as has been studied in \cite{Tobing19a},
optimization with only spectral features reconstruction does not yield sufficient conversion performance.
It has been found that with a cycle-consistent approach, i.e., cyclic VAE (CycleVAE),
significant improvements have been observed for the accuracy of converted speech. During the
optimization of a CycleVAE, converted spectral features that are generated with the speaker code of the
desired target speaker are used to generate cyclically reconstructed spectra that can also be optimized.

To synthesize a converted speech waveform from converted acoustic features generated from a VC model,
two different approaches can be used, i.e., using a conventional vocoder, such as STRAIGHT \cite{Kawahara99} or WORLD \cite{Morise16},
and using a neural vocoder, such as WaveNet \cite{Oord16,Tamamori17}. The latter is capable of producing natural-quality
speech in a copy-synthesis procedure and natural-to-high-quality
speech in text-to-speech \cite{Shen18} or VC \cite{Liu18} systems. In this work, instead of using an autoregressive (AR)
neural vocoder, such as WaveNet, we use a non-AR neural vocoder that generates waveform samples in parallel,
which will be more convenient for user-friendly baseline implementation. In particular, we use the recent non-AR neural
vocoder based on a generative adversarial network (GAN), i.e., Parallel WaveGAN (PWG) \cite{Yamamoto20}, which provides
real-time decoding with GPU and is able to generate high-quality output.

\begin{figure*}[t!]
  \centering
  \includegraphics[height=0.25\textheight]{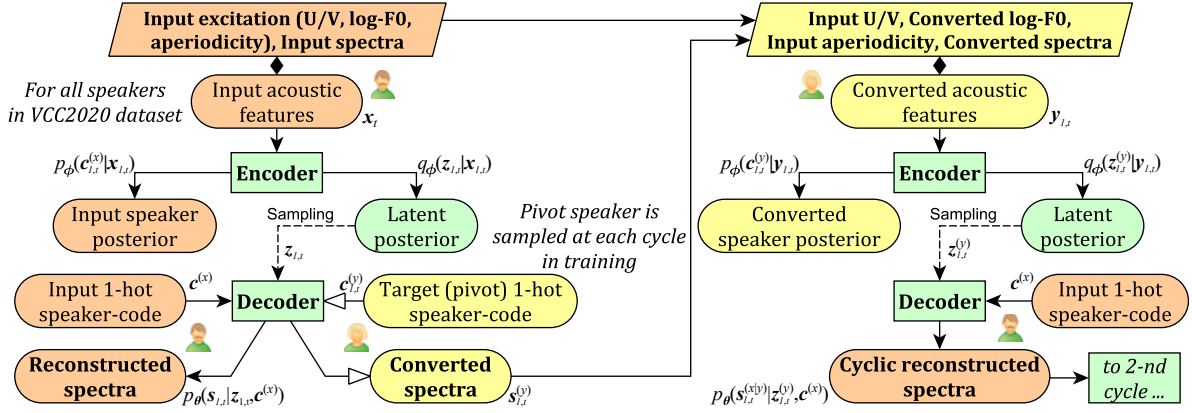}
  \vspace*{-4.5mm}
  \caption{Process flow of CycleVAE-based spectral modeling within the 1$^{\text{st}}$ cycle, where converted acoustic features of
  a sampled target (pivot) speaker are used to generate cyclically reconstructed spectra that can also be utilized during optimization.
  (pale-orange color refers to input-speaker part; pale-yellow color refers to pivot-/target-speaker part; green color refers to speaker-normalized part)}
  \label{fig:cycvae}
  \vspace*{-6mm}
\end{figure*}

To obtain better converted speech waveforms, we further reduce the mismatches \cite{Tobing19b} between the typical
natural acoustic features used to train a neural vocoder and the converted acoustic features generated from a VC model.
Specifically, we obtain reconstructed and cyclically reconstructed spectra from a CycleVAE-based spectral model,
and use them alongside the natural spectra to train a PWG-based neural vocoder. In practice,
the CycleVAE and PWG (CycleVAEPWG) baseline system can be developed in a straightforward manner using
our open-source implementation\footnotemark \hspace{0.25em}with the VCC 2020 dataset.
\footnotetext{\scriptsize {\url{https://github.com/bigpon/vcc20_baseline_cyclevae}}}

\vspace*{-0.5mm}
\section{CycleVAE-based Spectral Modeling}
\vspace*{-0.75mm}
\subsection{Model formulation}
\label{ssec:cycvae_formula}
\vspace*{-1.25mm}

Let $\vec{x}_t=[\vec{s}^{(x)^{\top}}_t,\vec{e}^{(x)^{\top}}_t]^{\top}$ be an input acoustic feature vector,
where $\vec{s}^{(x)}_t$ and $\vec{e}^{(x)}_t$ respectively denote an input spectral feature vector, such as that of mel-cepstrum
parameters, and an input excitation feature vector, such as that of fundamental frequency (F0), unvoiced/voiced (U/V) decision, and aperiodicity,
at time $t$. Likewise, the target (pivot) acoustic feature vector is denoted as $\vec{y}_t=[\vec{s}^{(y)^{\top}}_t,\vec{e}^{(y)^{\top}}_t]^{\top}$
and the input acoustic feature vector conditioned on the pivot speaker is denoted
as $\vec{x}^{(y)}_t=[\vec{s}^{(x|y)^{\top}}_t,\vec{e}^{(x|y)^{\top}}_t]^{\top}$ at time $t$.

As described in \cite{Tobing19a,Tobing20}, the variational lower bound for the CycleVAE model at the $n$-th cycle is given by
\begin{align}
    &\mathcal{L}(\vec{\theta}\!,\vec{\phi};\vec{x}_{n,t},\vec{x}_{n,t}^{(y)}) \!=\!
        \mathbb{E}_{q_{\vec{\phi}}(\vec{z}_{n,t}|\vec{x}_{n,t})}[\log p_{\vec{\theta}}(\vec{s}^{(x)}_{n,t}\!=\!\vec{s}^{(x)}_t|\vec{z}_{n,t},\vec{c}^{(x)})] \nonumber \\
        &\:+\mathbb{E}_{q_{\vec{\phi}}(\vec{z}^{(y)}_{n,t}|\vec{y}_{n,t})}[\log p_{\vec{\theta}}(\vec{s}^{(x|y)}_{n,t}=\vec{s}^{(x)}_t|\vec{z}^{(y)}_{n,t},\vec{c}^{(x)})] \nonumber \\
        &\:-\!D_{\text{KL}}(q_{\vec{\phi}}(\vec{z}_{n,t}|\vec{x}_{n,t})||p_{\vec{\theta}}(\vec{z}_{n,t}))
            \!-\!D_{\text{KL}}(q_{\vec{\phi}}(\vec{z}^{(y)}_{n,t}|\vec{y}_{n,t})||p_{\vec{\theta}}(\vec{z}^{(y)}_{n,t})) \nonumber \\
        &\:+\log p_{\vec{\phi}}(\vec{c}^{(x)}_{n,t}\!=\!\vec{c}^{(x)}|\vec{x}_{n,t})\!+\!\log p_{\vec{\phi}}(\vec{c}^{(y)}_{n,t}\!=\!\vec{c}^{(y)}_{n}|\vec{y}_{n,t}),\!\!\!\!
\label{eq:vlb}
\end{align}
where
\begin{align}
    \vec{s}^{(x)}_{n,t} &= g_{\vec{\theta}}(\vec{z}_{n,t},\vec{c}^{(x)}); \vec{s}^{(x|y)}_{n,t} = g_{\vec{\theta}}(\vec{z}^{(y)}_{n,t},\vec{c}^{(x)}), \label{eq:rec_cycrec} \\
    \vec{z}_{n,t} &= f_{\vec{\phi}}(\vec{x}_{n,t})^{(\vec{\mu})} - f_{\vec{\phi}}(\vec{x}_{n,t})^{(\vec{\sigma})} \odot \vec{\epsilon}, \\
    \vec{z}^{(y)}_{n,t} &= f_{\vec{\phi}}(\vec{y}_{n,t})^{(\vec{\mu})} - f_{\vec{\phi}}(\vec{y}_{n,t})^{(\vec{\sigma})} \odot \vec{\epsilon}, \\
    \vec{\epsilon} &= \sign(\vec{U})\ln(1\!-\!2|\vec{U}|), \,\text{s.t.}\: \vec{U}\!\!\sim\!(-1/2,1/2], \\
    \vec{x}_{n,t} &= [\vec{s}^{(x|y)^{\top}}_{n-1,t},\vec{e}^{(x)^{\top}}_{t}]^{\top}; \vec{s}^{(x|y)}_{0,t} = \vec{s}^{(x)}_t \\
    \vec{y}_{n,t} &= [\vec{s}^{(y)^{\top}}_{n,t},\vec{e}^{(y)^{\top}}_{n,t}]^{\top}; \vec{s}^{(y)}_{n,t} = g_{\vec{\theta}}(\vec{z}_{n,t},\vec{c}^{(y)}_{n}), \\
    \vec{c}^{(x)}_{n,t} &= f_{\vec{\phi}}(\vec{x}_{n,t})^{(\vec{c})}; \vec{c}^{(y)}_{n,t} = f_{\vec{\phi}}(\vec{y}_{n,t})^{(\vec{c})}, \\
    p_{\vec{\theta}}(\vec{z}_{n,t})&=p_{\vec{\theta}}(\vec{z}^{(y)}_{n,t})=\mathcal{L}(\vec{0},\vec{I}); \vec{c}^{(y)}_{n} \in \vec{C} \setminus \vec{c}^{(x)}.
\label{eq:vlb_aux}
\end{align}
Here, the time-invariant speaker code of the input speaker is denoted as $\vec{c}^{(x)}$ and that of the pivot speaker is denoted as $\vec{c}^{(y)}_{n}$. The
set of all available speaker codes is denoted as $\vec{C}$.

The set of parameters of the inference (encoder) network and that of the generative (decoder) network are respectively denoted as $\vec{\phi}$ and $\vec{\theta}$.
The feedforward functions of the encoder and decoder are respectively denoted
as $f_{\vec{\phi}}(\cdot)$ and $g_{\vec{\theta}}(\cdot)$. The latent feature vectors are denoted as $\vec{z}_{n,t}$ and $\vec{z}_{n,t}^{(y)}$.
The standard Laplacian distribution is denoted as $\mathcal{L}(\vec{0},\vec{I})$. The encoder network outputs
the parameters of latent posterior distribution, i.e., location ($\vec{\mu}$) and scale ($\vec{\sigma}$), and the logits of speaker posterior ($\vec{c}$).
The excitation feature vector of the sampled target speaker at the $n$-th cycle $\vec{e}^{(y)}_{n,t}$ consists of the transformed F0 \cite{Toda07},
input U/V decisions, and the input aperiodicity of $\vec{e}^{(x)}_t$.

\begin{figure*}[t!]
  \centering
  \includegraphics[height=0.25\textheight]{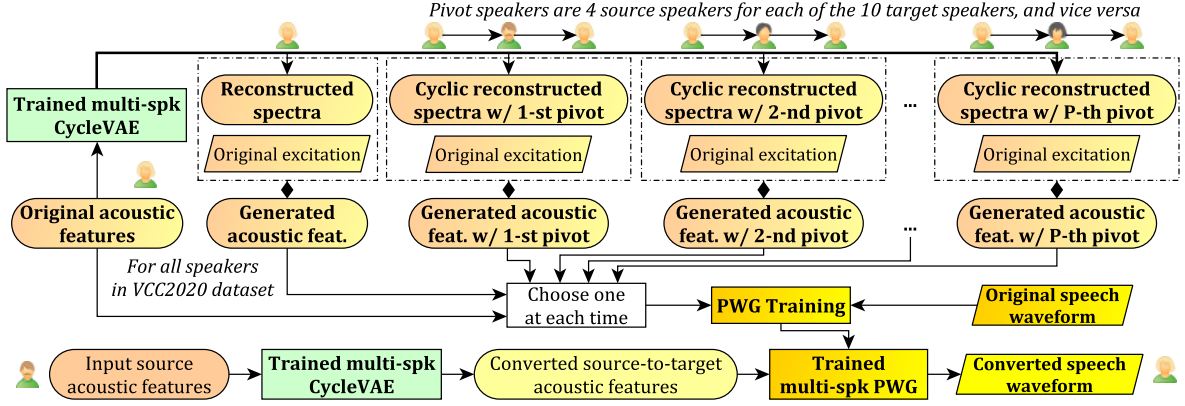}
  \vspace*{-2.5mm}
  \caption{Process flow for data augmentation with generated acoustic features from CycleVAE-based spectral model to train
  PWG-based neural vocoder, which is used to generate converted speech from the converted acoustic features in the conversion phase.
  (pale-orange color refers to source-speaker features part; pale-yellow color refers to target-speaker features part;
  gradient pale-orange-yellow color refers to the features set of all speakers; gradient orange-yellow color refers to the waveform set of all speakers)}
  \label{fig:pwg_aug}
  \vspace*{-4.5mm}
\end{figure*}

\subsection{Network architecture and training procedure}
\vspace*{-1mm}

An overview of the process flow for the CycleVAE-based spectral model, as described in Eqs.~\eqref{eq:vlb}--\eqref{eq:vlb_aux}, is shown in Fig.~\ref{fig:cycvae}.
The CycleVAE network architecture \cite{Tobing19a} consists of convolutional input layers and a recurrent layer with one
fully connected output layer, where its output is fed back to the recurrent layer. This structure applies to both encoder and
decoder networks. The extraction of acoustic features requires the annotation of the F0 range and power threshold values for each speaker, as described in \cite{Kobayashi18b}.
During optimization, a power threshold is used to automatically discard starting and ending silence frames. A loss function based on mel-cepstral distortion \cite{Toda07}
is used for the conditional probability density function (pdf) terms of the spectral features in Eq.~\eqref{eq:vlb}, whereas a cross-entropy loss function is used
for the corresponding conditional pdf terms of the speaker codes. In this baseline system, only the VCC 2020 dataset is used to develop the CycleVAE-based spectral model,
as well as the PWG-based neural vocoder.

\section{PWG-based Neural Vocoder}
\vspace*{-0.5mm}
\subsection{Network architecture and optimization}
\label{sseq:pwg_net-formula}
\vspace*{-1mm}

Parallel WaveGAN (PWG) \cite{Yamamoto20} is a GAN-based neural vocoder composed of a generator (G) network and a discriminator (D) network. The generator network
is designed to be similar to the WaveNet architecture conditioned on auxiliary acoustic features, where the output layer directly generates waveform samples
in parallel from input noise. The generator learns the waveform distribution by trying to deceive the discriminator in identifying the generated waveform
samples as the real ones, i.e., adversarial training. The discriminator tries to identify the original waveform samples as the real ones and the generated waveform
samples as the fake ones. To improve the modeling of waveform samples, a set of multiresolution short-time Fourier transform
(STFT) losses is also further included.

\subsection{Training procedure with data augmentation}
\label{ssec:pwg_aug}
\vspace*{-1mm}

In this system, we perform data augmentation for the acoustic features to obtain a more robust PWG-based neural vocoder with respect to the
mismatches \cite{Tobing19b} between naturally extracted spectral features $\vec{s}^{(x)}_t$ and converted spectral features $\vec{s}^{(y)}_{1,t}$,
which are generated from the CycleVAE model as given in Section~\ref{ssec:cycvae_formula}. Specifically, reconstructured spectral features
$\vec{s}^{(x)}_{1,t}$ and cyclically reconstructed spectral features $\vec{s}^{(x|y)}_{1,t}$ are used alongside the natural spectral
features $\vec{s}^{(x)}_t$ to train the PWG, as shown in Fig.~\ref{fig:pwg_aug}.

Therefore, the optimization for PWG can be written as
\begin{align}
    &\{\hat{\vec{\vartheta}},\hat{\vec{\varphi}}\} = \argmin_{\vec{\vartheta}}\argmax_{\vec{\varphi}}
        \mathbb{E}_{\vec{h}}[(1-D_{\vec{\varphi}}(G_{\vec{\vartheta}}(\vec{h}|\vec{x})))^2] \nonumber \\
        &+\!\!\mathbb{E}_{\vec{h}}[(1\!-\!D_{\vec{\varphi}}(G_{\vec{\vartheta}}(\vec{h}|\hat{\vec{x}})))^2]
            \!+\!\sum_{p=1}^{P}\!\mathbb{E}_{\vec{h}}[(1\!-\!D_{\vec{\varphi}}(G_{\vec{\vartheta}}(\vec{n}|\hat{\vec{x}}^{(y_{p})})))^2] \nonumber \\
        &+\!\!\mathbb{E}_{\vec{w}}[D_{\vec{\varphi}}(\vec{w})^2]
            \!+\!\mathbb{E}_{\vec{h},\vec{w}}[\text{STFT}_{\!m}\!(\vec{w},\!\hat{\vec{w}})]
                \!+\!\mathbb{E}_{\vec{h}\!,\vec{w}}[\text{STFT}_{\!m}\!(\vec{w}\!,\!\hat{\vec{w}}^{(x)})] \nonumber \\
        &+\!\!\sum_{p=1}^{P}\!\mathbb{E}_{\vec{h}\!,\vec{w}\!}[\text{STFT}_{\!m}\!(\vec{w}\!,\!\hat{\vec{w}}^{(x|y_{p})})]\,\text{s.t}\,\vec{h}\!\sim\!\!\mathcal{N}(0,\!I),\!\vec{w}\!\!\sim\!\!p(\vec{w}),\!\!\!
\label{eq:obj_gan_pwg_aug}
\end{align}
where $\vec{h}=[h_1,\dotsc,h_T]^{\top}$ denotes the sequence of white noise and the standard Gaussian distribution is denoted as $\mathcal{N}(0,I)$.
The parameter sets of the generator and discriminator are denoted as $\vec{\vartheta}$ and $\vec{\varphi}$, respectively.
The sequences of real and generated waveform
samples are denoted as $\vec{w}=[w_1,\dotsc,w_T]^{\top}$ and $\hat{\vec{w}}=[\hat{w}_1,\dotsc,\hat{w}_T]^{\top}$, respectively, where $\hat{\vec{w}}=G_{\vec{\vartheta}}(\vec{h}|\vec{x})$.
The distribution of the real waveform samples is denoted as $p(\vec{w})$. $\text{STFT}_{\!m}(\cdot)$ denotes $m$ sets of multiresolution STFT losses as in \cite{Yamamoto20}.
$\hat{\vec{x}}=[\hat{\vec{x}}_1^{\top},\dotsc,\hat{\vec{x}}_T^{\top}]^{\top}$ denotes the sequence of reconstructed acoustic features
with $\hat{\vec{x}}_t=[\vec{s}_{1,t}^{(x)},\vec{e}^{(x)}_t]^{\top}$ and $\hat{\vec{x}}^{(y_{p})}=[\hat{\vec{x}}^{(y_{p})^{\top}}_1,\dotsc,\hat{\vec{x}}^{(y_{p})^{\top}}_T]^{\top}$
denotes the sequence of cyclically reconstructed acoustic features with $\hat{\vec{x}}^{(y_{p})}_t=[\vec{s}_{1,t}^{(x|y_{p})},\vec{e}^{(x)}_t]^{\top}$.
Hence, $\hat{\vec{w}}^{(x)}=G_{\vec{\vartheta}}(\vec{h}|\hat{\vec{x}}^{(x)})$ and $\hat{\vec{w}}^{(x|y_p)}=G_{\vec{\vartheta}}(\vec{h}|\hat{\vec{x}}^{(x|y_p)})$.
For each speaker $x$, the number of pivot speakers $p$ to generate cyclically reconstructed spectra $\vec{s}_{1,t}^{(x|y_p))}$ in Eq.~\eqref{eq:rec_cycrec} is $P$.

\section{Results of VCC 2020}
\vspace*{-0.5mm}

\subsection{CycleVAEPWG baseline system conditions}
\vspace*{-1mm}

We used WORLD \cite{Morise16} to perform acoustic feature extraction. As the spectral features, $49$-dimensional mel-cepstrum
parameters, including the $0$-th power coefficient, were extracted from the speech spectral envelope with $0.466$ frequency warping coefficient.
CheapTrick \cite{Morise15} was used to extract the spectral envelope, Harvest \cite{Morise17} was used to extract F0 values,
and D4C \cite{Morise16b} was used to extract the aperiodicity spectrum, which was interpolated into $3$-dimensional code aperiodicity.
The sampling rate of the speech signal was $24000~$Hz. The number of FFT points in the analysis was $2048$. The frame shift was set to $5~$ms.

The VCC 2020 dataset \cite{vcc2020} consisted of 14 speakers, where 4 of them were the source speakers and the
remaining 10 were the target speakers. The set of source speakers consisted of 2 male and 2 female English speakers.
The set of target speakers consisted of another 4 English speakers (2 males and 2 females),
2 German, 2 Finnish, and 2 Mandarin speakers, where the latter three sets consisted of 1 male and 1 female.
The number of utterances for each speaker was 70 in the speaker's language. The English sets
had 20 parallel utterances between the source and target speakers. The sets of each of the other languages
were parallel. To train both CycleVAE and PWG, 60 utterances from each speaker were used in the training
set, whereas the remaining 10 utterances were used in the validation set. The evaluation set in the official test
consisted of another 25 utterances in English. Task 1 of VCC 2020 was intralingual conversion, whereas
Task 2 was cross-lingual conversion. The CycleVAEPWG baseline system was developed using only the VCC 2020 dataset.

\begin{figure}[t!]
  \centering
  \includegraphics[width=\linewidth]{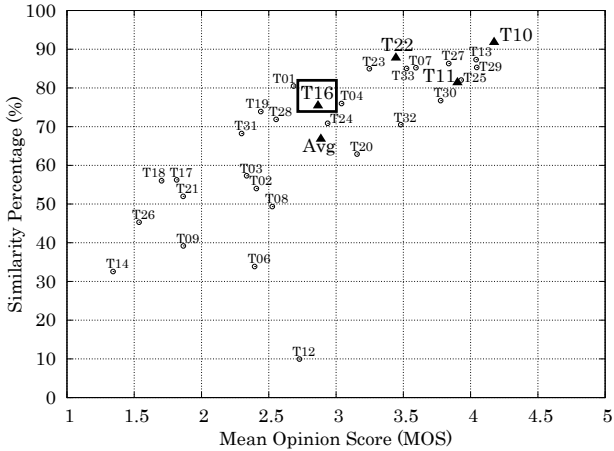}
  \vspace*{-4mm}
  \caption{Evaluation result of VCC 2020 Task 1 (intralingual conversion) consisting of MOS for naturalness
  ($x$-axis) and speaker-similarity percentage ($y$-axis), where the CycleVAEPWG baseline is denoted as T16, the other
  baseline systems are denoted as T11 and T22, and the top system is denoted as T10. The values are the
  average of the results of Japanese and English listeners. The average of all systems
  (31 including 3 baseline systems) is denoted as Avg.}
  \label{fig:task1}
  \vspace*{-5mm}
\end{figure}

Our CycleVAEPWG implementation is freely available at \url{https://github.com/bigpon/vcc20_baseline_cyclevae}.
The hyperparameters of CycleVAE were those in \cite{Tobing19a}. The differences were that the CycleVAE of VCC 2020 baseline system was implemented
for many-to-many VC on a unified encoder--decoder using standard Laplacian prior and only 2 cycles
as in \cite{Tobing20}. The CycleVAE spectral model was trained for $\sim\!\!2$ days with NVIDIA Titan V.

The hyperparameters of PWG were those in \cite{Yamamoto20} with the same configurations for STFT losses. The difference was
that the conditioning acoustic features consisted of mel-cepstrum parameters and the same
excitation parameters as in \cite{Wu20}. Furthermore, the generator of PWG was trained for 100K steps without a discriminator
and adversarial losses, i.e., only STFT losses, and another 300K steps with a discriminator and adversarial losses.
For the data augmentation procedure described in Section~\ref{ssec:pwg_aug}, each of the 4 source speakers had the 10 target speakers
as the pivot speakers $p$, and vice versa. The PWG neural vocoder was trained for $\sim\!\!3$ days with NVIDIA Titan V.

\subsection{VCC 2020 evaluation conditions}
\vspace*{-1mm}

In the VCC 2020 evaluation \cite{vcc2020}, a large-scale listening test was conducted, where
there were 206 Japanese listeners and 68 English listeners. There were three
baseline systems including CycleVAEPWG for both Tasks 1 and 2. The
total number of participant systems including the three baseline systems
for Task 1 was 31, whereas that for Task 2 was 28. The listening test
consisted of 1) naturalness evaluation, where each listener was asked to
evaluate the naturalness of each audio sample using a 5-scale
mean opinion score (MOS), and 2) speaker similarity evaluation, where each
listener was presented with two audio samples (converted and reference) and asked to
evaluate whether the two samples were produced by the same speaker using a 4-scale similarity score.
In the evaluation results, the baseline CycleVAEPWG is denoted
as T16, whereas the other two baseline systems are denoted as T11 \cite{Liu18} and T22 \cite{Huang2020}.
The top system in the VCC 2020 is denoted as T10.

\begin{figure}[t!]
  \centering
  \includegraphics[width=\linewidth]{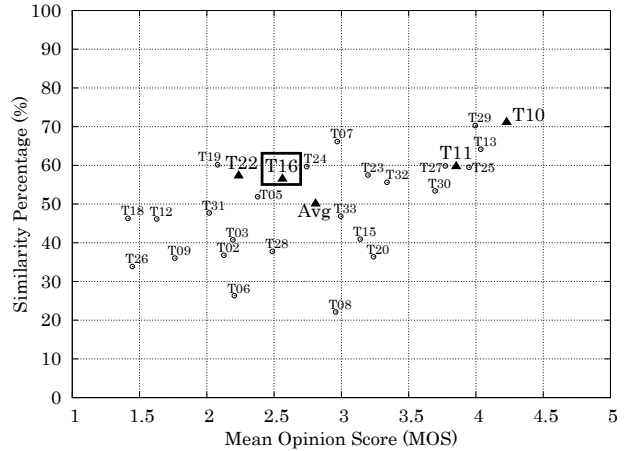}
  \vspace*{-4mm}
  \caption{Evaluation result of VCC 2020 Task 2 (cross-lingual conversion) consisting of MOS for naturalness
  ($x$-axis) and speaker-similarity percentage ($y$-axis), where the CycleVAEPWG baseline is denoted as T16, the other
  baseline systems are denoted as T11 and T22, and the top system is denoted as T10. The values are the
  average of the results of Japanese and English listeners. The average of all systems
  (28 including 3 baseline systems) is denoted as Avg.}
  \label{fig:task2}
  \vspace*{-5mm}
\end{figure}

\subsection{Evaluation results of Tasks 1 and 2}
\vspace*{-1mm}

The result of Task 1 (intralingual) is shown in Fig.~\ref{fig:task1} and the result
of Task 2 (cross-lingual) is shown in Fig.~\ref{fig:task2}, where the average
score of all participant systems including the baseline systems is denoted as Avg.
The values are the average results of Japanese and English listeners.
It can be observed that for Tasks 1 and 2, the CycleVAEPWG baseline achieves similarity percentages
of 75.37\% and 56.46\%, which are above the average similarity percentages of 66.76\% and 50.03\%,
respectively. As for naturalness, the CycleVAEPWG baseline achieves MOSs of 2.87 and 2.56 for Tasks 1 and 2,
respectively, which are about the same as the average MOS for Task 1 and below the average MOS for Task 2,
i.e., 2.89 and 2.81, respectively. Note that one other team, T23 \cite{nuvcc20}, also uses the CycleVAE-based
spectral model, but with the AR neural vocoder, i.e., WaveNet, for Task 2, which yields a significantly higher MOS of 3.20.

\vspace*{-1.15mm}
\section{Conclusions}
\vspace*{-1mm}

We have presented the description of CycleVAEPWG baseline system for the VCC 2020. The CycleVAEPWG
baseline consists of the cyclic variational autoencoder (CycleVAE)-based spectral model and
Parallel WaveGAN (PWG)-based neural vocoder, which is developed with only the VCC 2020 dataset using
freely available software. In the training, CycleVAE optimizes reconstructed and
cyclically reconstructed spectra, where the latter is obtained by recycling converted acoustic features.
The results of VCC 2020 have demonstrated that the CycleVAEPWG baseline 1) achieves an average MOS of
approximately 2.87 for naturalness for Task 1 (intralingual) and a below average MOS of 2.56 for Task 2 (cross-lingual),
2) achieves above average speaker similarity percentages of 75.37\% and 56.46\% for Tasks 1 and 2,
respectively. This system merits further research because of its solid performance for
intra- and cross-lingual VC, and its clear-cut implementation, which is freely available.

\vspace*{-1.15mm}
\section{Acknowledgements}
\vspace*{-1mm}
This work was partly supported by JSPS KAKENHI Grant Number 17H06101 and JST, CREST Grant Number JPMJCR19A3.

\bibliographystyle{IEEEtran}

\bibliography{vc20_cyclevaepwg}

\end{document}